\documentclass[useAMS,usenatbib]{mn2e_modif}
\usepackage{graphicx,epsfig}
\usepackage{amssymb}
\newcommand{\tcc}{2008~TC$_3$}
\newcommand{\tc}{TC$_3$}
\newcommand{\as}{Almahata Sitta}

\title[On the origin of the Almahata-Sitta meteorite and \tcc\ asteroid]
{On the origin of the Almahata-Sitta meteorite and \tcc\ asteroid}

\author[J. Gayon-Markt et al.]  {Julie Gayon-Markt$^1$\thanks{E-mail:
    julie.gayon@oca.eu}, Marco Delbo$^1$, Alessandro Morbidelli$^1$, Simone Marchi$^{1,2}$\\
  $^1$Laboratoire Lagrange UMR 7293, Universit\'e Nice Sophia-Antipolis, CNRS, Observatoire de la C\^ote d'Azur,
  B.P. 4229, 06304 Nice Cedex 4, France.\\
  $^2$NASA Lunar Science Institute, Center for Lunar Origin and Evolution, Southwest Research Institute,
  1050 Walnut St, Suite 300, Boulder, CO 80302, USA.  }

\begin{document}

\date{Accepted 2012 May 1.}

\maketitle

\begin{abstract}
  Asteroid \tcc\ was a Near Earth Asteroid that impacted the
  Earth on 2008 October 7. Meteorites were
  produced by the break-up of \tcc\ in the high atmosphere and at
  present, about $600$ meteorites - called Almahata Sitta - coming
  from \tcc\ have been recovered.  A mineralogical
  study of Almahata Sitta fragments shows that the asteroid \tcc\ was
  made of meteorites of different types (ureilites, H, L, and E
  chondrites). Understanding the origin of this body and how it was
  put together remain a challenge.  Here we perform a detailed
  spectroscopical and dynamical investigation to show that the most
  likely source region of \tcc\ is in the inner Main Belt at low
  inclination ($i<8^\circ$). We show that asteroids with spectroscopic
  classes that can be associated with the different meteorite types of
  \as\ are present in the region of the Main Belt that includes the
  Nysa-Polana family and objects of the Background at low inclination.
  Searching for a possible scenario of formation for \tcc, we show
  that there is little chance that \tcc\ was formed by low velocity
  collisions between asteroids of different mineralogies, in the
  current asteroid belt. It seems more likely that the heterogeneous
  composition of \tcc\ was a inherited from a time when the asteroid
  belt was in a different dynamical state, most likely in the very
  early Solar System. Because ureilites are fragments of a large,
  thermally metamorphosed asteroid, this suggests that the phases of
  collisional erosion (the break-up of the ureilite parent-body) and
  collisional accretion (the formation of the parent body of \tcc)
  overlapped for some time in the primordial asteroid belt.

\end{abstract}

\begin{keywords}
techniques: spectroscopic --
catalogues --
meteorites, meteors, meteoroids --
minor planets, asteroids: individual: \tcc 
\end{keywords}

\section{Introduction}
\label{section_intro}

Meteorites are a partial sample of asteroids that survive the passage
through the Earth's atmosphere. The identification of the source
regions of the different type of meteorites is essential to be able to
link the mineralogical properties of meteorites to the parent
asteroids and, consequently, to address the mineralogical evolution in
the asteroid belt.  However, this is not an easy task and only some of
these links could be established: for instance, the group of
howardites, eucrites, and diogenites (HEDs) meteorites are thought to
come from the Vesta family of asteroids \citep[e.g.][]{binzel93}; more
speculatively, L ordinary chondrites could come from the Gefion family
\citep{nesvorny09}, while asteroids of the the Flora family bear
spectral similarities with the LL chondrites
\citep{vernazza08}. However, the parent bodies of most meteorite
types, if still intact, are unknown.

The discovery and spectroscopic observation of the near-Earth asteroid
(NEA) \tcc\ (henceforth \tc) 20 hours before it impacted the Earth's
high atmosphere, and the subsequent recovery of meteorites (called
Almahata Sitta) -- clearly coming from this body -- was a major result
in this respect \citep{jenniskens09, shaddad10}. It allowed a direct
link between an asteroid and meteorites to be established for the
first time: the asteroid was classified as belonging to the
spectroscopic F-class \citep[in the Tholen
classification,][]{tholen84} or B-class \citep[in the Bus
classification,][]{busbinzel02} on the basis of the flat shape of its
reflectance spectrum in the region between 500 and 1000 nm.  Moreover,
among the 47 meteorites initially recovered, it was observed that the
visible spectrum of the fragment $\#7$ matches the telescopic spectrum
of \tc\ obtained before the impact with the Earth's atmosphere
\citep{jenniskens09}. Fragment $\#7$ is an achondrite polymict
ureilite \citep{jenniskens09}. Ureilites are in a group of achondritic
meteorits that exhibit both primitive and evolved characteristics
\citep{cloutis10}: in particular, they are characterized by olivine
and pyroxene-rich clasts among carboneous material (mainly graphite);
fine-grained graphite is also present, which lower the albedo of the
meteorites \citep[about 10-12\%;][]{hiroi10}. Ureilites were initially
thought to derive from $S$-class asteroids (see for instance
\citealt{gaffey93}).  However, because of its spectral properties,
\citet{jenniskens09} propose a link with $B$-class asteroids according
to the Bus classification or $F$-class in the Tholen
classification. This is more plausible than a link with $S$-class
asteroids, given the low albedos of ureilite meteorites, consistent
with $B$ or $F$-class asteroids but not with the $S$-class.

It is worth to note that the F-class can be distinguished from the
B-class from a much weaker UV drop-off in the spectra of the former
compared to the latter and also because B-class asteroids show a
moderately higher average albedo than F-class bodies.  However, in the
Bus classification these two classes are merged in a unique class (the
B-class).  This is because the \citet{busbinzel02} spectral
classification is based on spectra acquired with CCD spectrographs,
which -- in general -- do not observe far enough in the UV to observe
the above-mentioned drop-off feature.  We will refer in this work to
B-class asteroids including both Tholen B- and F-classes, and Bus
B-class.

Interestingly, mineralogical studies of \as\ show that, among the
$\sim 600$ fragments recovered \citep{shaddad10}, about $70-80\%$ are
ureilites, while the remaining $20-30\%$ are enstatite chondrites, H
and L ordinary chondrites.  More specifically, \citet{bischoff10} show
that, among a subsample of 40 deeply studied meteorites from \as, 23
fragments are achondritic ureilites and 17 have chondritic litologies
with 14 of them corresponding to enstatite chondrites, 2 to H ordinary
chondrites and 1 to a new type of chondrite (see \citealt{horstmann10}
for more details). Although small clasts of different types are quite
common in brecciated meteorites \citep[see][for a
review]{meibomclark99} and carboneous material is found in some HED
meteorites, it is the first time that meteorites of very different
mineralogies (i.e primitive fragments with achondrite polymict
ureilites and evolved ones such as ordinary chondrites or enstatite
chondrites) are associated to the same fall. This led to make the
hypothesis that \tc\ was an asteroid made of blocks of different
mineralogies, held together very loosely \citep[given the explosion of
the body at the anomalously high
altitude;][]{jenniskens09,bischoff10,shaddad10}.

Tracing back \tc\ to its source region in the asteroid Main Belt would
allow us to understand the origin of the Almahata Sitta meteorites and
how \tc\ was put together by loosely assembling meteorites of
different mineralogies. Establishing this link would also be
fundamental to shed light on the source region of ureilites, that
albeit rare, are the forth major class of primitive meteorites
recovered on Earth after the CV, CI and CO carbonaceous chondrites.

In their attempt to find the source region of \tc\ and \as,
\citet{jenniskens10} selected all $B$-class asteroids, according to
the Bus classification, and objects of the Tholen F and B classes and
searched for spectra similarities with \tc\ and \as. As a result of
their study, they showed spectral similarities between \tc\ and
ungrouped asteroids as well as several dynamical asteroid groups (or
families) as possible origins for the \tc\ asteroid, namely Polana
($2.4$ AU, $3^\circ$), Hoffmeister ($2.8$ AU, $4.5^\circ$), Pallas
($2.8$ AU, $33^\circ$), Themis ($3.15$ AU, $1.5^\circ$), and Theobalda
($3.2$ AU, $14^\circ$). Later, from dynamical grounds,
\citet{jenniskens10} concluded that asteroids from the inner asteroid
belt (i.e. with orbital semimajor axis $a<2.5$AU) are the likely
parent bodies of \tc. This reduces the choice to dispersed B-class
asteroids in the inner main belt and the Polana asteroid group.  In
Section~\ref{section_dynamics}, we revisit this issue studying the
possible dynamical source regions for \tc.

We recall here that the Polana group is part of a cluster of asteroids
known as the Nysa-Polana family \citep{nesvorny10}, which is located
in the inner main belt, between the $\nu_6$ secular resonance (at
heliocentric distance $\approx$ 2.1 AU) and the 3:1 mean motion
resonance with Jupiter (at heliocentric distance of 2.5 AU). This
family has a complex -- twofold -- structure in orbital proper element
space \citep{nesvorny10}, suggesting that it is the outcome of at
least two independent break-up events in the same orbital region. From
the few spectral data available at the time, \citet{cellino01} argued
that the Nysa-Polana family contains asteroids of three spectral
classes. The first class is that of B-class objects, like asteroid
(142) Polana itself -- note that \citet{cellino01} uses the F-class
classification from the \cite{tholen84} taxonomy; the second class is
the S-class, with the largest member being identified as the asteroid
(878) Mildred; the third class is that of X-class objects, such as the
asteroid (44) Nysa.  In this manuscript, we revisit this result using
a much wider dataset of spectro-photometric data provided by the
Moving Objects Catalog of the Sloan Digital Survey
\citep[SDSS,][]{ivezic02}, which is analyzed here using a new
classification algorithm \citep[][described in Section
\ref{section_method}]{michel05} developed for the Gaia space mission
of the European Space Agency.

A detailed study of the mineralogy of the Nysa-Polana family is of
great importance also for better understanding the origin of other
NEOs. In partcular, \cite{campins10} claimed that the asteroid
(101955) 1999 RQ$_{36}$, target of the sample return mission
OSIRIS-REx (approved in the program New Frontiers of NASA), was
delivered to near-Earth space via the $\nu_6$ secular resonance from
the Polana group. Moreover, the binary asteroid (175706) 1996 FG$_3$,
primary target of the sample return mission Marco Polo-R, under study
at the European Space Mission (ESA), might have formed within the
Polana group and delivered to the near-Earth space via the overlapping
Jupiter 7:2 and Mars 5:9 mean motion resonances rather than the
$\nu_6$ \citep[see][]{walsh12}.

As a consequence, in Sections~\ref{section_polana} and
\ref{background}, we perform a spectroscopic analysis using the SDSS
data of the asteroids of the Nysa-Polana family as well as dispersed
asteroids of the inner Main Belt (called objects of the background) in
order to find spectral matches with \tc\ and \as.

Finally, in Section~\ref{section_formation}, we investigate a possible
formation scenario for the \tc\ asteroid as a rubble pile of rocks of
different mineralogical types, which is based on the peculiar low
inclination of the Nysa-Polana family and dispersed asteroids.

\section{Dynamical history and main-belt origin of \tc}
\label{section_dynamics}

The large majority of NEAs are fragments generated by the collisional
disruption of larger asteroids of the main belt; said fragments drift
in orbital semimajor axis by the so-called Yarkovsky effect until they
reach regions of orbital instabilities (mean motion resonances with
Jupiter and secular resonances) which, by enhancing their orbital
eccentricities, deliver them to the near-Earth space \citep[see][for a
review]{morby02}.

According to model by \citet{bottke02}, there are five main
intermediate sources of NEAs:
\begin{itemize}
\item[1)] the $\nu_{6}$ secular resonance, which marks the inner edge
  of the asteroid belt and occurs when the precession frequency of the
  longitude of perihelion of an asteroid is equal to that of Saturn;
\item[2)] the 3:1 mean motion resonance located at $a~\sim 2.5$AU,
  where the orbital period of an asteroid is 1/3 of that of Jupiter;
\item[3)] Mars-Crossing asteroids, defined as objects which are not
  NEAs (i.e. their perihelion distance $q$ is larger than 1.3 AU), but
  whose semimajor axis evolves in a random-walk fashion as a result of
  close and distant encounters with Mars.
\item[4)] the outer belt population, whose eccentricities can increase
  up to planet-crossing values due to a network of high order orbital
  resonances with Jupiter and three-body resonance of type
  asteroid-Jupiter-Saturn \citep{morbynesvorny99}
\item[5)] dormant Jupiter family comets.
\end{itemize}

The orbital elements of \tc\ before impact are not known very
precisely. Here are different estimates: $a = 1.29$ AU, $e = 0.299$,
and $i = 2.441^\circ$ (NEODyS website:
http://newton.dm.unipi.it/neodys); $a = 1.308$ AU, $e = 0.312$, and
$i= 2.542^\circ$ \citep{jenniskens10}.  For each of these orbits we
computed, using the original orbital evolution files of
\citet{bottke02}, which are the most likely intermediate sources among
those enumerated above. We found that \tc\ has a probability of
63-66\% to come from the $\nu_{6}$ source and 34-37\% to come from the
Mars crosser population. Apparently, none of the simulations for the
3/1 resonance in \citet{bottke02} produce objects within $\pm .05$ AU
in $a$, $\pm .05$ in $e$ and $\pm 2.5^\circ$ in $i$ of \tc. This
contrasts with the claim reported in \citet{jenniskens10}, that \tc\
has a 20\% probability to come from the 3/1 resonance in the
\citet{bottke02} model. We suspect that there has been an error in the
manipulation of the \citet{bottke02} model in that work.

Of the Mars crossers that can produce the \tc\ orbit in
\citet{bottke02} simulations, none has semimajor axis larger than 2.5
AU. From this, and remembering that the $\nu_{6}$ resonance lays at
the inner edge of the belt, we conclude - in agreement with
\citet{jenniskens10} - that \tc\ most likely comes from the inner
asteroid belt, inside of 2.5 AU.

As noted in \citet{jenniskens10}, there are two populations of
asteroids in the inner belt with spectra broadly consistent with that
of \tc: the Polana members, with an orbital inclination of 2--3
degrees, and a population of dispersed B-class asteroids, with
inclinations ranging up to 15 degrees. These objects are too dispersed
to belong to a relatively young collisional family, but their broad
cluster in orbital space suggests that they might belong to an old
collisional family, dynamically dispersed probably during the phase
when the orbits of the giant planets changed substantially, about 4
Gyr ago.

To determine whether it is more likely that \tc\ comes from the
Nysa-Polana family or from the population of dispersed B-class
asteroids, we turn, once again, to the original simulations in
\citet{bottke02}.  In that work, the initial distribution of asteroids
in the $\nu_{6}$ was uniform in inclination. In particular, 50\% of
the initial conditions had initial inclination larger than 8
degrees. However, 93\% of the particles which reproduced the orbit of
\tc\ at some time during their evolution, have initial inclination
$i<8^\circ$. For the Mars crossers, all those reproducing the orbit of
\tc\ have initial $i<8^\circ$. This suggests that the most likely
source of \tc\ is at low inclination, consistent with the Nysa-Polana
family and with the dispersed B-class asteroids with $i<8^\circ$. This
is not surprising, given that the inclination of \tc\ is 2.3--2.5
degrees. However, the dynamics rules out a \tc\ origin from the
dispersed B-class asteroids at higher inclination (i.e. i$>8^\circ$).

Finally, \citet{campins10} have shown that the Polana group can easily
deliver small enough fragments in the $\nu_6$ resonance. In fact, the
extrapolated Yarkovsky-induced semimajor axis distribution of Polana
group members predicts that asteroids fainter than $H\sim$ 18.5 can
reach the border of the $\nu_6$ resonance, which is at 2.15 AU for a
Polana-like inclination. Assuming a Polana-like geometric visible
albedo of 0.05, $H=$ 18.5 translates into a diameter $D\sim$ 2 km,
which is much larger then the size of \tc.

For all these reasons we conclude that, from the dynamical viewpoint,
the Nysa-Polana family and dispersed B-class asteroids with
$i<8^\circ$ are the most likely sources of \tc. Thus, we will focus on
the Nysa-Polana family and the Background at low inclination in the
next sections.

\section{Spectroscopic analysis method}
\label{section_method}

\subsection{Asteroid spectral classification algorithm}
\label{subsection_algo}
In order to search for spectral groups in the Nysa-Polana asteroid
family or in the inner main belt background at low inclination, we
used the unsupervised classification algorithm that will be adopted
for the classification of asteroid spectra from the multiband
photometers \citep{jordi10} on board the Gaia space mission of the
European Space Agency (ESA).

While Gaia is mainly devoted to the observations of $10^9$ stars, it
is expected that this mission will also observe more than 250,000
asteroids multiple times over 5 years (mission lifetime)
\citep{mignard07}. Spectral classification of asteroids will be
performed by an unsupervised classification algorithm based on the
works of \citet{michel05} and \citet{galluccio08}.  An unsupervised
approach has the advantage that no {\it a priori} information is taken
into account to build the spectral groups.

\begin{figure}
\centering
\includegraphics[width=\columnwidth,angle=0]{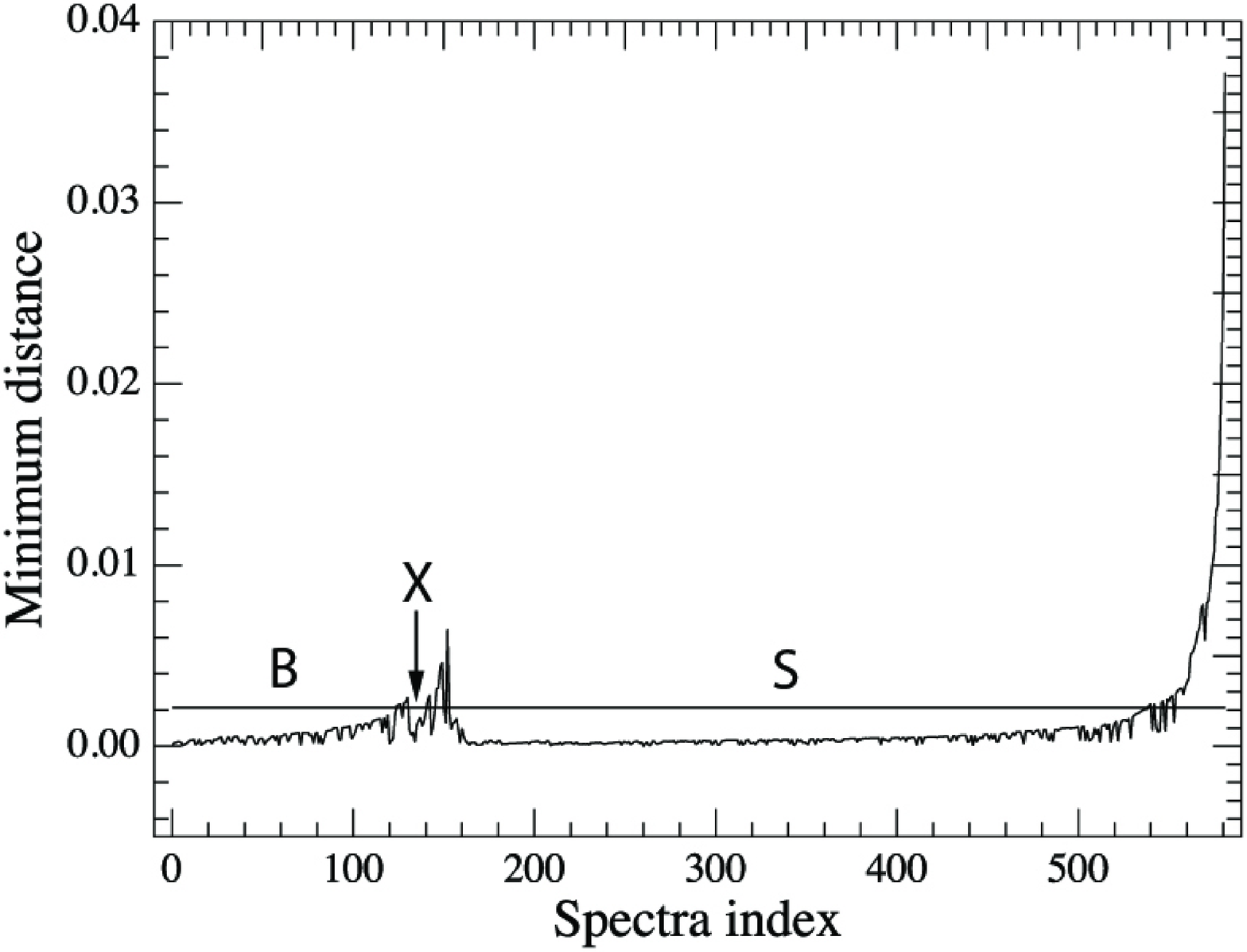}
\caption{Minimum distance between spectra computed by the MST method
  as a function of the spectra index.  The cut off used to obtain the
  B, X, and S groups was determined such as already classified
  asteroids belong to their appropriate group.  }\label{Fig:MST}
\end{figure}

The algorithm is based on a method for partitioning a set $V$ of $N$
data points ($V \in \mathbb{R}^L$) where $L$ in our case is the number
of spectral bands into $K$ non-overlapping clusters (or groups of data
points) such that (a) the inter-cluster variance is maximized and (b)
the intra-cluster variance is minimized. Each spectrum is considered
as a vertex of tree. A tree is a graph that is connected (i.e. every
vertex is connected to at least another vertex) and acyclic (i.e. the
graph has no loops). The spanning tree (MST, i.e. the tree passing
through each vertex of the set) with the minimal length is calculated
using the so-called Prim's algorithm \citep[see][and references
therein]{galluccio09}. The length of each edge of the tree (i.e. the
distance between two spectra) is determined using the Kullback-Leibler
metric: let $v_i=\{v_{i1},\ldots,v_{iL}\}$ the feature vector
corresponding to a reflectance spectrum ($L$ being the number of
wavelengths); at a given wavelength $\lambda_j$, each spectrum is
associated with a (positive) normalized quantity:
$\tilde{v_{ij}}=v_{ij}/\sum_{j=1}^L v_{ij}$, which can be interpreted
as the probability distribution that a certain amount of information
has been measured around the wavelength $\lambda$. To measure the
similarity between two probability density functions, we compute the
symmetrized Kullback-Leibler divergence:
\[d_{KL}(v_i||v_k) = \sum_{j=1}^{L} (\tilde{v_{ij}}-{\tilde v_{kj}})\log 
\frac{ \tilde{v_{ij}}}{ \tilde{v_{kj}}}.\]\

The identification of the cluster is performed by first computing, at
each step of the MST construction, the length of the newly connected
edge and then by identifying valleys in the curve obtained by plotting
the MST edge length as a function of the iteration of the construction
(see Figure \ref{Fig:MST}).  The valleys in this curve identify the
number and the position of high density region of points, i.e. the
clusters \citep[see][for a more thorough description of the
algorithm]{galluccio09}.

\subsection{Spectral data}
\label{subsection_data}
In order to increase the sample statistics compared to the relatively
small number of asteroids observed by \cite{cellino01} and identify
meaningful spectral groups within the Nysa-Polana family, we used our
spectral classification algorithm on the observations contained in the
SDSS MOC4 \citep[][http://www.sdss.org]{ivezic02}. The MOC4 contains
magnitudes of 104449 Main Belt asteroids.  Each asteroid was observed,
in general, at multiple epochs over 5 spectral bands in the visible
light, namely $u'$, $g'$, $r'$, $i'$, $z'$ at the following central
wavelengths ($\lambda$) of $354$, $477$, $623$, $763$, $913$ nm,
respectively.

To obtain the surface reflectance of the asteroids, we needed to
remove the solar contribution ($u_\odot$, $g_\odot$, $r_\odot$,
$i_\odot$, $z_\odot$) to the observed magnitudes.  The solar
contribution is calculated from transformation equations between the
SDSS u'g'r'i'z' magnitudes and the usual $UBVR_cI_c$ system. We find:
$u_\odot=6.55$, $g_\odot=5.12$, $r_\odot=4.68$, $i_\odot=4.57$,
$z_\odot=4.54$.  (See the SDSS website for more details:
http://www.sdss.org).  In order to compare asteroid spectral
reflectances and be able to classify them, we choosed to normalize
each spectrum relatively to the $r'$ band ($\lambda_{r'}=623$
nm). Finally, we obtained the following asteroid reflectance for the
$\lambda_{u'}$ wavelength:
\[F_{u'} = 10^{-0.4 (C_{u'} - C_{r'}) }\]
\[\textrm{with }C_{u'} = u' - u_\odot \textrm{ and } C_{r'} = r' - r_\odot\] 
the reflectance colors at $\lambda_{u'}$ and $\lambda_{r'}$
respectively.  The same computations were performed for the
$\lambda_{g'}$, $\lambda_{i'}$, and $\lambda_{z'}$ wavelengths.

We searched in the MOC4 database for all asteroids belonging to the
Nysa-Polana family and for dispersed objects.  Concerning the family
members identification, we used the definition of dynamical families
of \citet{nesvorny10}.  Among the Nysa-Polana family, 4134 objects
have been observed at least once by SDSS (as a comparison, the number
of spectra obtained in the visible light or near-infra red, for the
Nysa-Polana family, in the SMASS, ECAS and 52-Color catalogs are : 15,
13, and 2, respectively).  For those asteroids with more than one
observation, we calculated the weighted mean reflectance by averaging
the reflectancies derived at each epoch weighing each point with its
respective signal-to-noise ratio.

However, likely due to non-photometric conditions some of the observed
objects have high uncertainties in the measured magnitudes. In the
visible light, the separation of asteroid spectroscopic group is based
on the overall slope of their reflectance and in the presence (or
absence) and strength of the 1 \micron\ absorption feature. In order
to avoid the superposition of spectral groups we selected asteroid
with the less noisy observations. Namely, we rejected observations
with a relative uncertainty $>10\%$ on the in-band photometric flux
derived from the MOC4 magnitudes.  In the end, the taxonomic
classification of the Nysa-Polana dynamical family was performed over
$579$ objects and over $2828$ objects for the Background at low
inclination.

\begin{figure}
  \centering     
     \includegraphics[scale=0.32,angle=270]{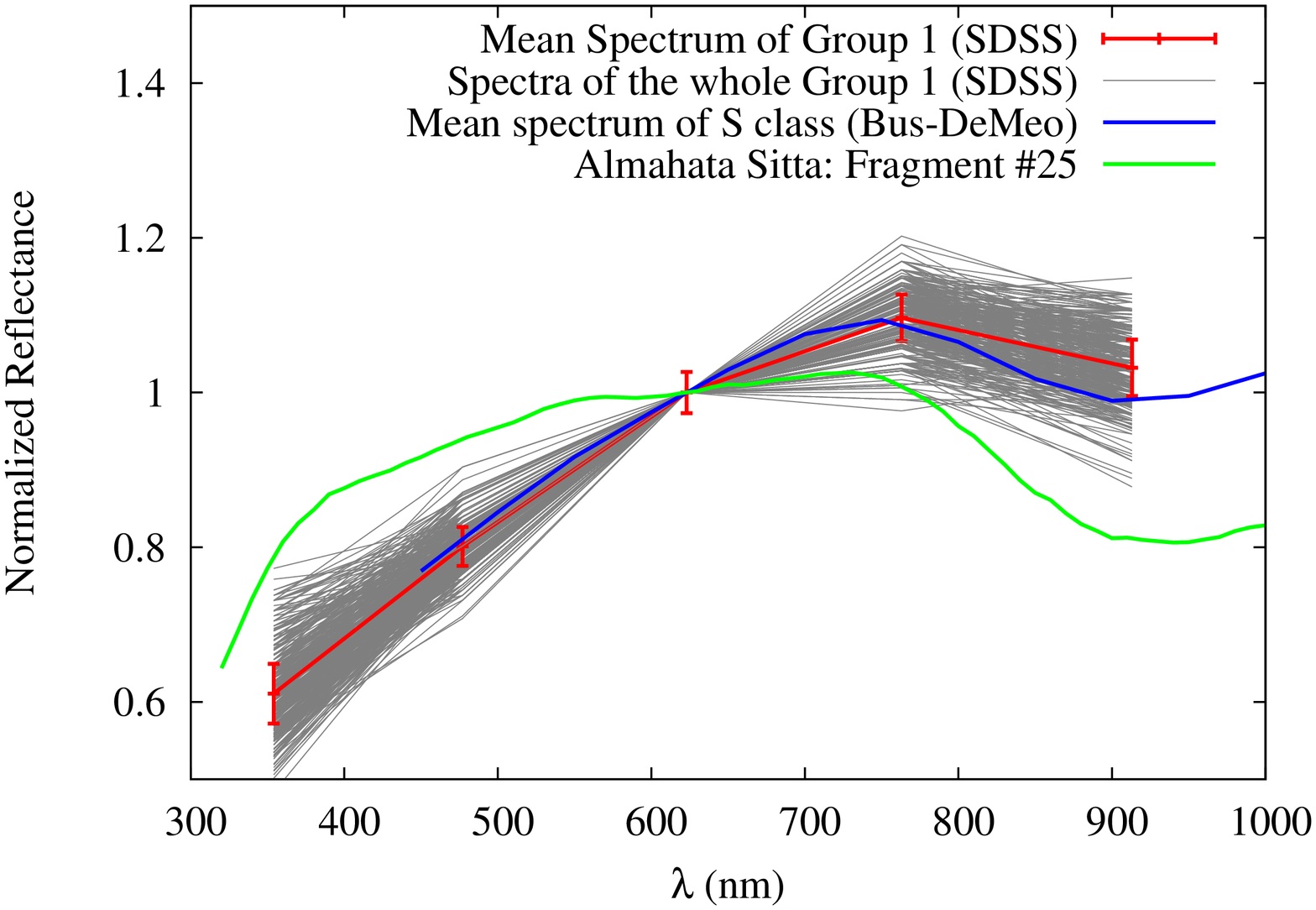}\\
     \includegraphics[scale=0.32,angle=270]{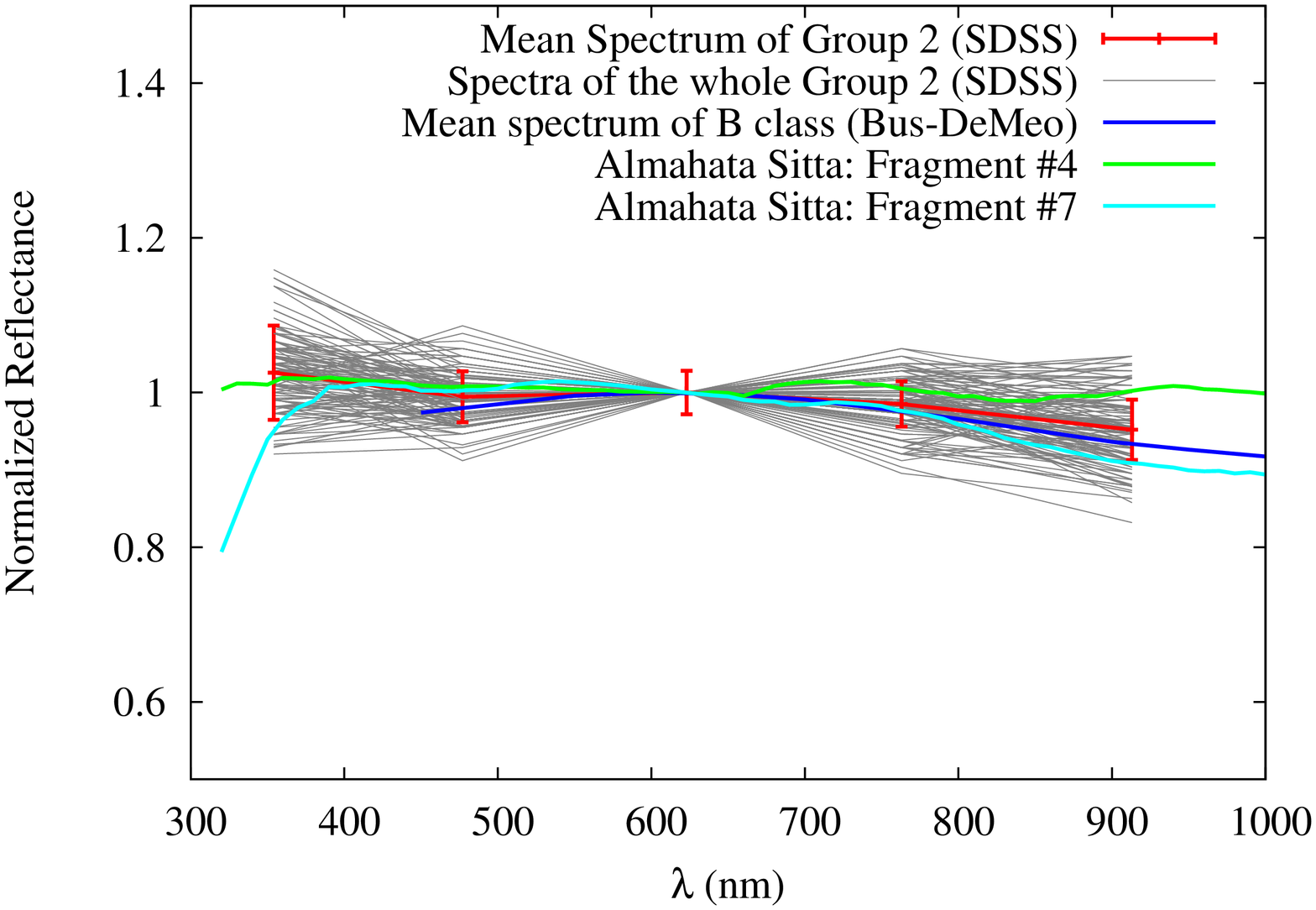}\\
     \includegraphics[scale=0.32,angle=270]{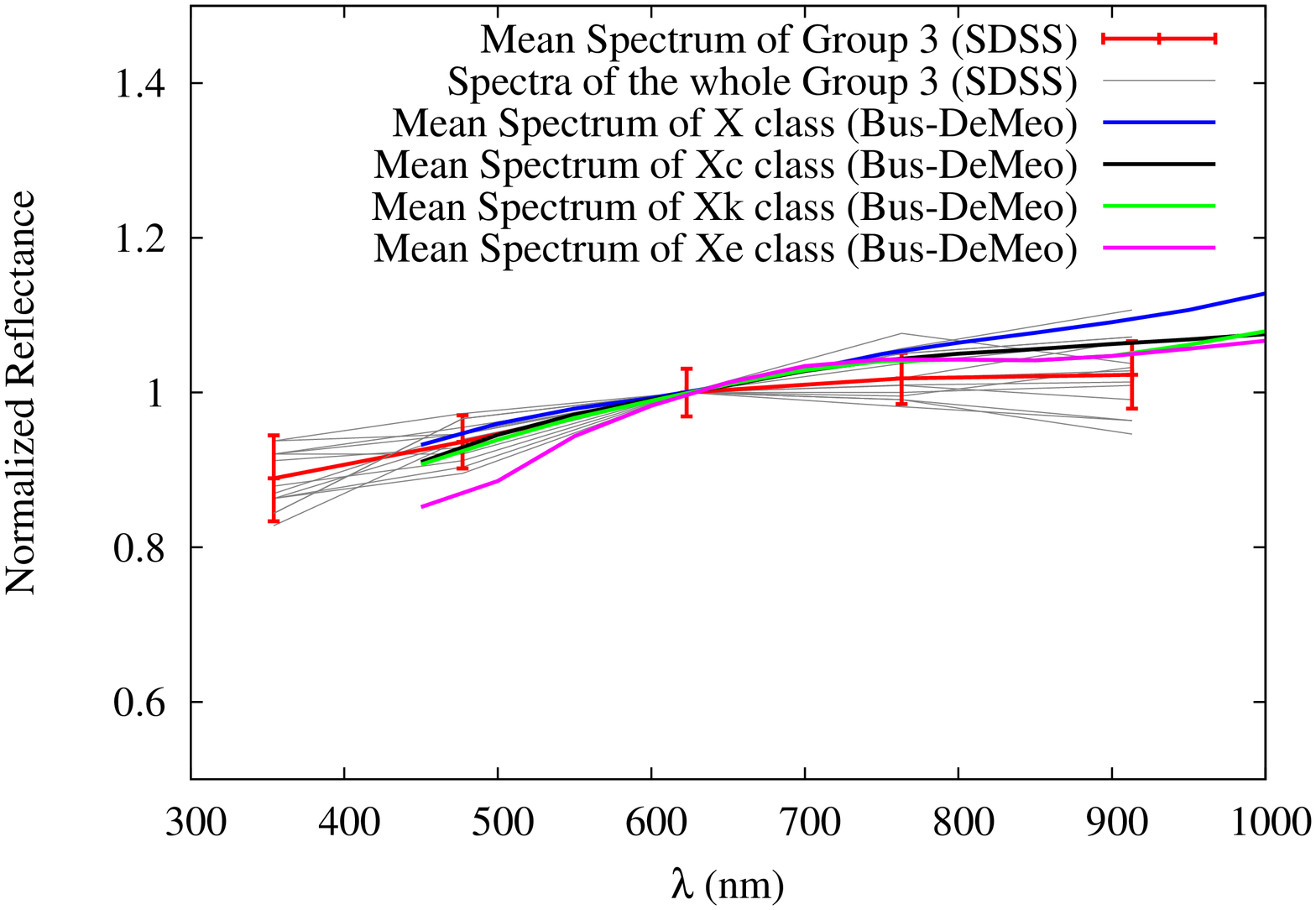}
     \caption{Panel (a), (b), and (c) represent the spectra for each
       group with their mean spectrum.  A comparison with several
       spectra classes coming from the Bus taxonomy is also performed:
       Group~1 with $S$-class spectrum, Group~2 with $B$-class
       spectrum and Group~3 with $X$-subclasses. We also find a good
       agreement with several \as\ fragments.  Spectral differences
       with fragment $\#25$ (H-chondrite) appearing in panel (a) are
       likely due to space weathering.  Fragments $\#4$ and $\#7$ in
       panel (b) are ureilites.}
   \label{Fig:spectra} 
\end{figure}

\begin{figure}
  \centering     
     \includegraphics[scale=0.65,angle=0]{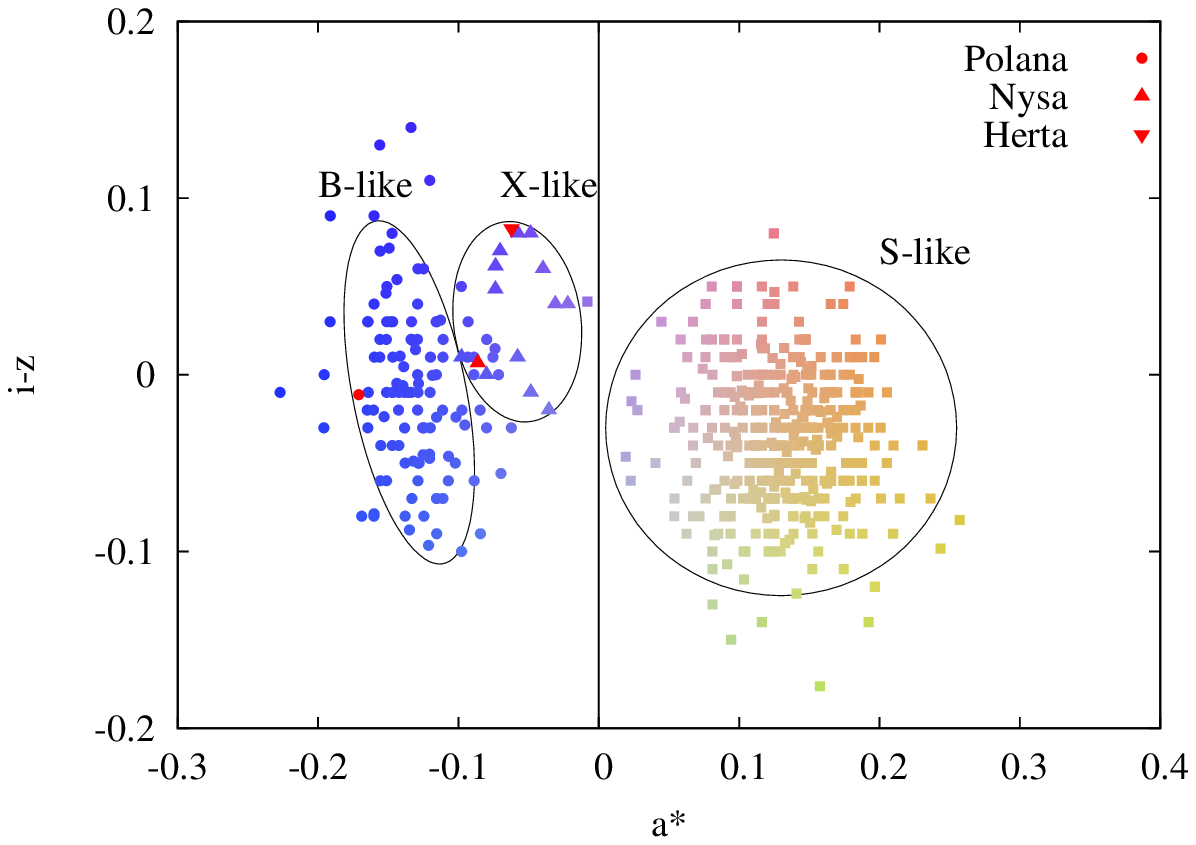} 
     \caption{Distribution of S-class ({\tiny $\blacksquare$}),
       B-class $\bullet$), and X-class ({\small $\blacktriangle$})
       objects of the Nysa-Polana family as a function of
       $a\textrm{*}$ and $i'-z'$ (calculated from SDSS
       magnitudes). Ellipses represent the location of S, B, and X
       regions.  The color palette is the same of \citet{parker08}.
       Although Polana (B), Nysa (X), and Herta (X) have not been
       observed by SDSS, we have calculated their $a\textrm{*}$ and
       $i'-z'$ values from the ECAS catalog and located them in this
       plot.}
 \label{Fig:astariz} 
\end{figure}

\section{Spectral analysis of the Nysa-Polana family}
\label{section_polana}

\subsection{Application of the spectral classification method}
\label{subsection_classification}
Figure \ref{Fig:MST} shows the MST length as a function of the number
of iterations (i.e. addition of spectra) in the case of the SDSS-MOC4
spectrophotometric data of the 579 asteroids of the Nysa-Polana
family.  The figure clearly shows three main valleys corresponding to
three groups composed of 118, 13, and 378 asteroids, respectively. The
remaining objects ($70$) are considered as unclassified asteroids.
This is due to their large value of the MST length which means no
spectral similarity among the family is found for these $70$ objects.

Figure \ref{Fig:spectra} shows asteroid spectra of each group and
their average spectrum. Error bars are the 1-$\sigma$ standard
deviation. We also compared the spectra of each spectral group of
asteroids with the mean spectrum of different taxonomic classes from
the Bus classification (http://smass.mit.edu/busdemeoclass.html).  In
each panel, we only plot the Bus mean spectra which present the closer
similarity with the mean spectra of each group:

\begin{enumerate}
\item For the first group (top panel of Figure \ref{Fig:spectra}, 378
  objects), we find that all spectra resemble the $S$-class mean
  spectra of Bus.
\item In the middle panel of Figure \ref{Fig:spectra}, the 118 spectra
  of the second group fits with $B$-class asteroids (such as 142
  Polana, for instance).
\item The bottom panel shows the third more populous group containing
  13 asteroids. Comparing the spectra of this group and the mean
  spectrum of the $X$-class, $X_c$-class, $X_k$-class, and
  $X_e$-class, we find that the third group of 13 objects corresponds
  to asteroids of the $X$-complex and its subclasses.
\end{enumerate}

We note that groups (1) and (2) are consistent with the two main
spectral classes found by \citet{cellino01}. Moreover, the same
authors had already identified the presence of asteroids with
$X$-class spectral reflectances. They also noted that two of the
largest asteroids, (44) Nysa and (135) Herta have spectra consistent
with an $X$-class. As a consequence, we confirm their results using
the SDSS MOC4 large sample of spectra.

In Figure \ref{Fig:astariz} we plot the asteroids of the three
spectroscopic groups of the Nysa-Polana family as a function of
$a\textrm{*}$ and $i'-z'$ (with $a\textrm{*}=0.89
(g'-r')+0.45(r'-i')-0.57$). We show that the $S$-class group is well
defined with $a\textrm{*} \in [0 ; 0.25]$ and $i'-z' \in [-0.15;
0.08]$.  The $B$-class region stretches from $\sim -0.2$ to $\sim
-0.06$ for $a\textrm{*}$ and from $-0.1$ to $0.14$ for $i'-z'$ and we
find for the $X$-class region, $a\textrm{*} \in [-0.1 ; -0.02]$ and
$i'-z' \in [-0.02 ; 0.08]$.  The color palette used in
Fig. \ref{Fig:astariz} is that of \cite{parker08}. Blue dots
correspond to asteroids with neutral or slightly blue spectra
(B-class), asteroids with neutral to slightly red spectra (X-class)
are displayed in the plot with purple colors, while S-class asteroids
are red to yellow.

\subsection{Comparison with Almahata Sitta meteorites}
\label{subsection_meteorites}
It is very interesting to note that the three asteroid spectroscopic
groups that we found in the Nysa-Polana family are likely to be the
analogs of the different meteorite mineralogies found in \as.  To
strengthen this argument, we also compare -- in Figure
\ref{Fig:spectra} -- the mean spectrum of each asteroid spectroscopic
class of the Nysa-Polana family (S, B, X; red curves) with published
spectra \citep[from][]{hiroi10} of some fragments of \as.  We note
that at the time of writing, spectra of E-chondrites from Almahata
Sitta are not yet publicly available. So, comparisons between the
classes of asteroids in the Nysa-Polana family and fragments of \as\
were performed for the S and B classes only.  However, comparison of
spectra of other E-chondrites \citep[from the RELAB database
from][]{gaffey76} with the average spectrum of the the Nysa-Polana
X-class shows a good agreement.

Concerning the B class, we find that the mean spectrum of the
Nysa-Polana B-class asteroids matches the spectra of the fragments
$\#4$ and $\#7$ (ureilites) of Almahata Sitta.  It is important to
note that this spectral match was obtained only considering the
visible wavelengths between about 350 and 900 nm.  We remind that,
from spectroscopic and albedo similarities,
\citet{jenniskens09,jenniskens10} have proposed a link between the
B-class (or the F-class) and ureilite meteorites.  An important caveat
is that the link between B-class asteroids and the \as\ ureilites
proposed by \citet{jenniskens09,jenniskens10} is based on the noisy
spectrum of \tc. Moreover, B-class asteroids are more commonly
associated with carbonaceous chondrites by several studies
\citep[see][for a review about B-class objects, their meteorite
analogues and their composition]{clark10}.  However, a recent
spectroscopic survey of B-class asteroids by \cite{deleon12} shows
that the ensemble of the reflectance spectra of the 45 B-class
asteroids analyzed in their work have a continuous shape variation in
the range between 500 and 2500 nm, from a monotonic negative (blue)
slope to a positive (red) slope. \cite{deleon12} apply a clustering
technique to reduce the ensemble of the spectra to 6 optimized
averaged spectra or ``groups''. Interestingly the RELAB spectrum of
the fragment \#4 of \as\ shows a good match with the group \#3 of
\cite{deleon12} in the region between 500 and 2500 nm (De Leon,
J. private communication September, 2011).

Concerning the S class, we show, in Fig. \ref{Fig:spectra}, that
fragment $\#25$ (an ordinary chondrite) is rather close to the S-class
mean spectrum, as expected, but not in very good agreement \citep[see
e.g. the review of][for spectroscopical links between ordinary
chondrites and S-class asteroids]{chapman04}. One possible explanation
for this spectral mismatch is the space weathering of S-class
asteroids. We discuss this possibility in the following subsection.

\subsection{Space weathering of S-class members of the Nysa-Polana family}
\label{subsection_spaceweathering}
Space weathering is a physical process caused by cosmic rays,
collisions, ion bombardment, that alters physical and spectral
properties of the surface of atmosphere-less planetary bodies. More
particularly, due to space weathering, S-class asteroids become darker
(their albedo is reduced), redder (the reflectance increases with
increasing wavelength), and the depth of absorption bands are
reduced. As a consequence, the spectral slope changes due to this
process \citep{clark02,chapman04}.

In \citet{marchi06}, a strong relation was found between the spectral
slope of S-class asteroids and their "exposure" to space
weathering. The exposure of an asteroid to space weathering
corresponds to time integral the flux of solar ions that the body
receives along its orbit.  The exposure to space weathering depends on
the age of the asteroid and its average distance from the
Sun. Therefore, in order to know if space weathering affected S-class
asteroids of the Nysa-Polana family, we calculate the exposure to
space weathering of these bodies as shown in \citet{marchi06} and
\citet{paolicchi07}.

\begin{figure}
  \centering
  \includegraphics[width=\columnwidth,angle=0]{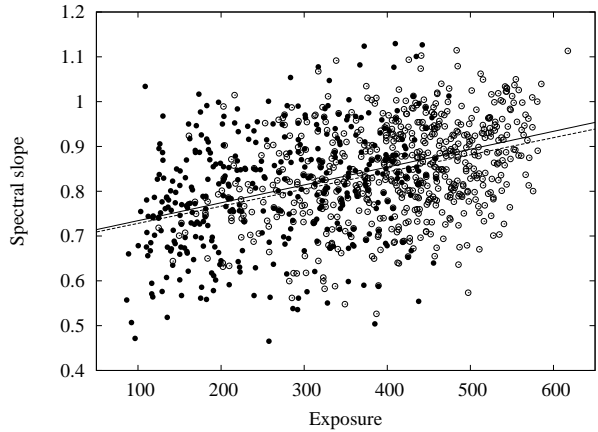}
  \caption{Spectral slope as a function of the exposure to space
    weathering of S-class objects of the Nysa-Polana family (black
    dots and black line) and S-class objects of the background at low
    inclination (circles and dashed line). As a comparison, the
    spectral slope of H-chondrites which are thought to come from
    S-class asteroids is of about $0.1-0.2$.}
   \label{Fig:spaceweathering} 
\end{figure}

\begin{figure*}
  \centering     
     \includegraphics[scale=0.32,angle=270]{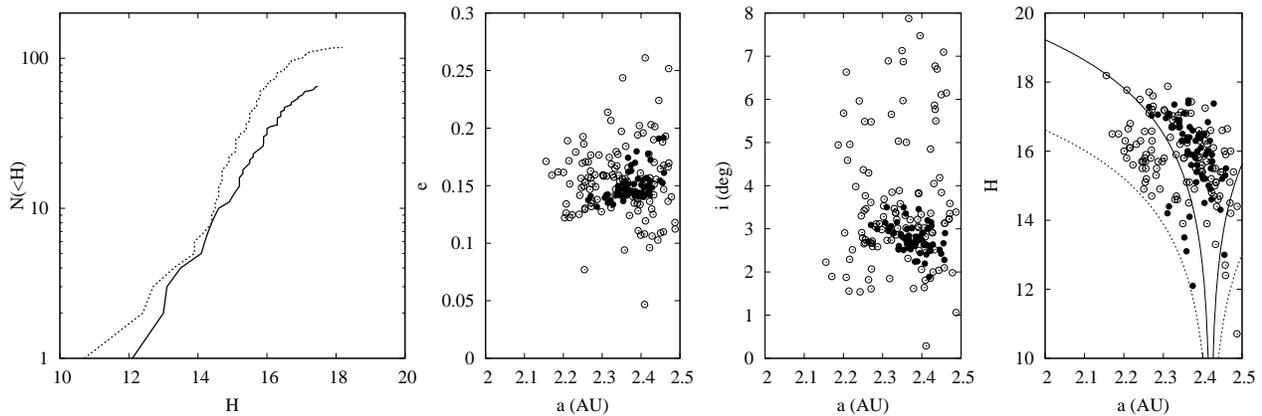}
     \caption{ Plot 1: Cumulative number of asteroid $N(<H)$ as a
       function of the $H$ absolute magnitude for B-class asteroids of
       the Nysa-Polana family (black curve) and the Background at low
       inclination ($i<8^\circ$; dashed curve).  Plot 2-4:
       Distribution in eccentricity (plot 2), inclination (plot 3) and
       $H$ absolute magnitude (plot 4) as a function of semimajor
       axis.  Black points correspond to Polana B-class asteroids and
       black circles to B-class dispersed objects from the Background
       at low inclination. Plot 4 shows the $H$-dependent semimajor
       axis distribution induced by the Yarkovsky effect that best
       fits the boundaries of the observed distribution for the Polana
       group (black curves) and for the Background at low inclination
       (dashed curve).  }
   \label{Fig:distribution} 
\end{figure*}

We compute the slope $\alpha$ of a SDSS spectrophotometric data by
fitting a linear equation normalized to $1$ at $\lambda_r$ ($623$ nm)
to the datapoints.  To calculate the exposure, we need to estimate the
average collisional age of Main Belt asteroids ($T_{MBA}$), which can
be constrained by the age of the Belt ($t_{LHB}\sim 4$ Gyr) and by
collisional times ($\tau_{coll}$):
\[T_{MBA} \simeq \tau_{coll} \left[ 1- \left( 1+
    \frac{t_{LHB}}{\tau_{coll}} \right)
  e^{-\frac{t_{LHB}}{\tau_{coll}}}\right] +
t_{LHB}\,e^{-\frac{t_{LHB}}{\tau_{coll}}}\] According to
\citet{bottke05}, the collisional lifetime can be estimated as a
function of asteroid diameter (using SDSS H absolute magnitudes and
assuming an albedo of $\rho_V = 0.2$ for all S-class objects).  In the
end, we have:
\[\textrm{exposure} \propto \frac{T_{MBA}}{a^2 \sqrt{1-e^2}}\]
with $a$ and $e$ the asteroid proper semimajor axis and eccentricity,
respectively.

In Figure \ref{Fig:spaceweathering}, we plot both the spectral slope
of Nysa-Polana S-class objects as a function of the exposure to space
weathering and the best linear fit to this distribution.  The
two-tailed probability for the linear correlation is lower than
$0.001\%$ which means the linear correlation is significant\footnote{A
  linear fit is considered significant whenever the two-tailes
  probability is lower then $5\%$.}.  This shows that the spectral
slope increases with the exposure.

Semimajor axes and eccentricities do not vary much within the
Nysa-Polana family.  As a consequence, the exposure variation mainly
comes from asteroid diameter, through the computation of $T_{MBA}$.
Hence, plotting the spectral slope as a function of the exposure, or
$T_{MBA}$, or asteroid diameters ($d$) provide quite similar figures
(not shown here) -- where the exposure increases with time or diameter
-- but with the following abscissa ranges: $0.5<T_{MBA}<2.5$ Gyr or
$1<d<6$ km.

The spectral slope increase with exposure or asteroid age is then the
proof that space weathering occurs for S-class asteroids of the
Nysa-Polana family.  We also note that the spectral slope of the
fragment $\#25$ of \as\ has the value of $0.03$.

As a consequence, the spectral mismatch between asteroids of the
S-component of the Nysa-Polana family and the ordinary chondrites of
\as\ -- as observed in Figure \ref{Fig:spectra} -- can be explained by
space weathering.

\section{Analysis of asteroids of the Background at low inclination}
\label{background}

As mentioned in Sections \ref{section_intro} and
\ref{section_dynamics}, the other possible origin source for \tc, from
a dynamical point of view, is the population of dispersed asteroids at
low inclination. In the present section, we perform the same
spectroscopic study as for the Nysa-Polana family (Section
\ref{section_polana}) but using objects of the Background located in
the Inner Main Belt ($2.1<a<2.5$ AU) with a proper inclination lower
than $8^\circ$. Selecting the best observations of the SDSS MOC4
catalog (i.e. observations with a relative uncertainty $>10\%$ on the
in-band photometric flux, corresponding to 2828 objects), we find 4
large groups of asteroids corresponding to the following spectral
classes: B, C, X, S, as well as small clusters of Q, V, and A-class
asteroids.  In addition, for the B, X, and S groups of the Background,
we obtain mean spectra very similar to those of the Nysa-Polana family
(not shown here). We also note that space weathering is also
found for S-class asteroids of the Background (see
Fig. \ref{Fig:spaceweathering}).  As a consequence, the three
ingredients required to form \tc\ (S-class asteroid/H-chondrite;
B-class asteroid/ureilite; X-class asteroid/enstatite chondrite) are
both found in the Nysa-Polana family and in the Background at low
inclination.

In our attempt to definitively conclude on the origin region of \tc,
we have compared the distribution of B-class asteroids for the two
source regions. In order to get a homogeneous selection of B-class
asteroids both from the Nysa-Polana family and the Background at low
inclination, we have selected objects such as their spectra are
similar to the B-class mean spectrum of the Nysa-Polana family, within
the 1-$\sigma$ error bar. Although this selection reduces the size of
the original B-class groups, we limit the overlap of classes (B and X
classes can overlap; see Fig. \ref{Fig:spectra}).

From this selection, we plot in Fig. \ref{Fig:distribution}, the
cumulative number of asteroids as a function of $H$ absolute magnitude
as well as the distribution in eccentricity, inclination and $H$
magnitude as a function of semimajor axis. We find more B-class
asteroids from the Background at low inclination (dashed curve in the
first plot of Fig. \ref{Fig:distribution}) than from the Polana group
(black curve). As a consequence, the Background at low inclination
seems to dominate the Polana group, even if, around magnitude 17, the
slope of the Polana group curve looks a little steeper than the curve
for the Background.  In the other plots in orbital elements, we can
see that the two regions overlap in some points and that B-class
objects of the Background are much more dispersed compared to Polana
asteroids.

As in \citet{campins10}, we also represent the absolute magnitude
$(H)$ a a function of semimajor axis, both for the Polana group and
the Background \citep[the plot of][was done for the whole Nysa-Polana
family]{campins10}.  The plot appears to be V-shaped which is a
feature known to be associated both with the size-dependent ejection
velocity field and with the drift in the proper semimajor axis induced
by the Yarkovsky effect \citep[e.g.][]{vokrouhlicky06}.  In the plot,
the black curve shows the $H$-dependent semimajor axis distribution
induced by the Yarkovsky effect that best fits the boundaries of the
observed distribution for the Polana B-class asteroids. Objects below
this curve are expected to be interlopers and may belong to the
Background. The extrapolated Yarkovsky-induced distribution predicts
that the Polana group should reach the outer edge of the $\nu_6$
resonance for objects with $H \sim 18$, which for a Polana-like albedo
of $p_v = 5\%$ translates into a diameter $D\sim2$ km. This means that
objects smaller than $2$ km, such as \tc, can easily escape the Polana
group and the Inner Main Belt.

Concerning the Background at low inclination, different size limits
can be computed due to a large variation of proper orbital
inclinations. For an inclination of $\sim0^\circ$, the $\nu_6$
resonance boundary is found at $a\sim2.1$ AU, which gives us an $H$
magnitude of $\sim 16$ and an asteroid diameter limit of $d\sim4$
km. For the highest inclination ($i\sim8^\circ$), we obtain $H \sim
14.5$ at $a\sim2.2$ AU, which is equivalent to a B-class asteroid
($p_v=5\%$) of 8 km. As a result, the Background at low inclination
could have also delivered \tc\ through the $\nu_6$ resonance.

Because the Background at low inclination is not a dynamical family,
its V-shaped structure in $H(a)$ was not especially expected. We then
think that the Background at low inclination could correspond to an
old break-up of the Nysa-Polana family.  As a consequence, we can
conclude that \tc\ comes from the Inner Main Belt, more particularly
from the Background at low inclination or the Nysa-Polana family, and
that these two sources could be genetically linked.

\section{Possible Formation Scenario of \tcc} 
\label{section_formation}

Meteorite strewn fields with fragments of different mineralogical
types are very rare. Almahata Sitta and Kaidun \citep{ivanov84} are
probably the only known specimen.  In the scenario that mixing occurs
by a collisional process, the paucity of mineralogically-mixed
meteorites suggests that the process that formed \as~i.e. mixing the
material between projectile and target is very rare; in most cases the
projectile is pulverized and leaves negligible traces in the target
\citep{melosh89}.

It is unclear which conditions allow for mixing between projectile and
target. Impact velocity is probably a key parameter. The average
impact velocity between asteroids in the main belt is 4.4-5.3 km/s
\citep{bottke94}.  If projectile/target mixing were possible at these
impact speeds, meteorites like Almahata Sitta would probably be
frequent, which is not the case. Thus, we think that unusually low
impact velocities are needed for mixing. This could prevent the target
from pulverizing and could lead to macroscopic projectile fragments
being implanted in the regolith of the impacted body or
gravitationally bound to fragments of the target, if the latter is
catastrophically disrupted by the impact.  The fact that we give
evidence that \tc\ comes from the Nysa-Polana family (see Section
\ref{section_polana}) or the Background at low inclination (see
Section \ref{background}), which are characterized by a mixing a
taxonomic classes, also suggests that a specificity of the members of
these families/regions may be unusually low collision velocities with
projectiles that are also on low-inclination orbits.

For all these reasons we did a systematic search for projectiles that
could hit Nysa-Polana family members or dispersed asteroids of the
Background at very low speeds.  We did this search using the algorithm
for the calculation of the intrinsic collision probability $P_i$
between pairs of asteroids, described in \citet{wetherill67}. This
algorithm, given the semimajor axis $a$, eccentricity $e$ and
inclination $i$ of the two orbits (we use proper values for each pair
of selected asteroids) assumes that the angles $M$, $\omega$, $\Omega$
(mean anomaly, argument of perihelion and longitude of the ascending
node) of the two objects have a uniform probability distribution over
the range $0 - 2\pi$; then it computes which fraction of these angles
corresponds to the two objects being closer to each other than 1 km;
finally, this fraction is translated into an intrinsic collision
probability per year ($P_i$), using the orbital periods of the two
objects and assuming that they are not in resonance with each other.
For our goals, we modified this algorithm in order to take into
account only orbital intersections corresponding to relative speeds
smaller than $0.5$ km/s. Admittedly, this velocity threshold is
arbitrary.  Given the exceptional character of \as, we need a
threshold much smaller than the typical impact velocity among random
asteroids, that is to say a sub-sonic impact velocity, in order to
preserve the target.

Because most of Almahata Sitta fragments recovered are mainly made of
ureilites, which we consider as analogs of B-class asteroids \citep[as
previously mentioned in Section \ref{section_intro} and according
to][]{jenniskens10}, we assume that the favorite scenario for the
formation of \tc\ involves a low velocity collision of asteroids of X
and S classes (projectiles) with a B-class member (target) of the
Nysa-Polana family.

\begin{figure}
  \centering     
  \includegraphics[width=\columnwidth,angle=0]{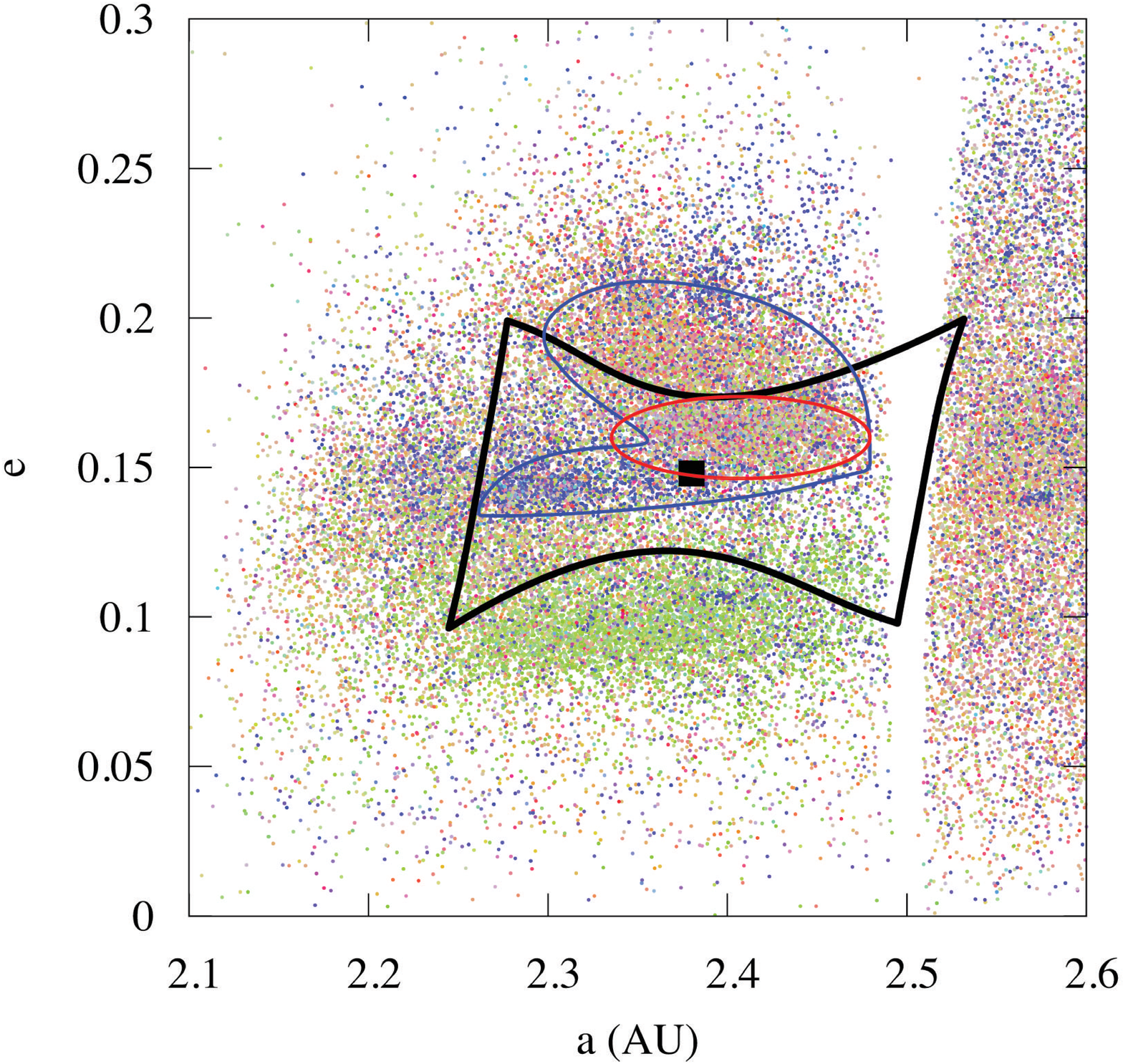} \\
  \includegraphics[width=\columnwidth,angle=0]{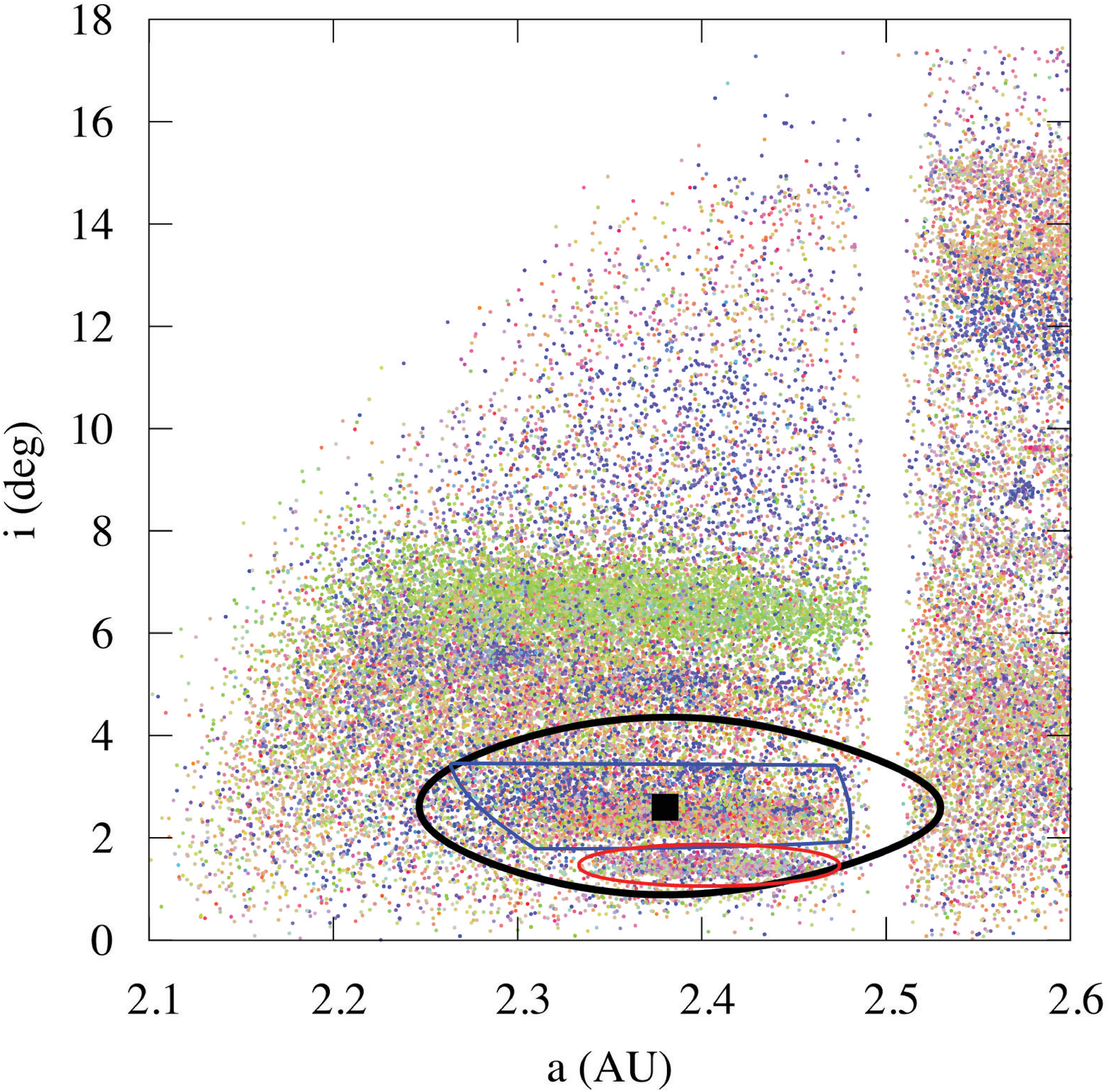}
  \caption{Distribution of proper orbital element for asteroids of the
    inner Main Belt: semimajor axes and eccentricities (top panel) or
    inclinations (bottom panel).  For an asteroid target (black
    square), the region of orbital element space where impactors can
    come from is plotted in black. The location of the Nysa-Polana and
    Massalia families are respectively represented in blue and red.
    The color palette corresponds to the Parker's color distribution
    as mentioned in \citet{parker08} and depends on the value of
    $a^{*}$ and $i'-z'$, with $a^{*}=0.89(g'-r')+0.45(r'-i')-0.57$,
    $g'$, $r'$, $i'$, $z'$ corresponding to SDSS magnitudes at the
    following central wavelengths: 477 nm, 623 nm, 763 nm and 913 nm,
    respectively.  Bluer dots ($a^{*}\in[-0.25 ; 0]$ and
    $i'-z'\in[-0.2 ; 0.2]$) are related to dark objects while yellow
    to red colors ($a^{*}\in[ 0 ; 0.25]$ and $i'-z'\in[ -0.2 ; 0.2]$)
    gather brighter objects (see Figure \ref{Fig:astariz} for more
    details).  Data come from the SDSS Moving Objects Catalog 4.}
   \label{Fig:collision} 
\end{figure}

As for projectiles we considered all asteroids known in the ASTORB
file (Asteroid Orbital Elements Database:
ftp://ftp.lowell.edu/pub/elgb/astorb.html) of the whole Main Belt and
we computed the probability of collision of each asteroid with a
B-class object near the center of the B-class group of the Nysa-Polana
family. Figure \ref{Fig:collision} shows the region of orbital element
space where impacts can occur (impact probability $>$0).  This figure
shows that asteroids included in the region of low-velocity impacts
have different SDSS colors and thus different taxonomies.  As a
consequence, collisions involving B-class, S-class, and X-class
asteroids are likely to be possible.

In the end, depending on the location of the target within the Polana
group, we find that projectiles come from one of the following
families: (i) Nysa-Polana, (ii) Flora or (iii) Massalia, located in
the inner Main Belt, or (iv) Hestia which is in the central Main Belt,
very close to the 3:1 mean motion resonance with Jupiter.  More
particularly, for targets with a semimajor axis ($a$) smaller or equal
than $2.3$ AU, the Flora family and objects of the background, very
close to the Flora and Nysa-Polana family are the main source of
projectiles; for targets with 2.3 $<a<$ 2.4 AU and eccentricities
larger than $\sim $0.15, projectiles in general are likely to come
from the Nysa-Polana family; for targets with 2.3 $<a<$ 2.4 AU and
eccentricities smaller than $\sim$ 0.15, projectiles are likely to
come from the Massalia family; for targets with $a\geq$ 2.4 AU
projectiles come from the Hestia family.  Similar results are found
when studying collisions involving B-class asteroids of the Background
at low inclination.

\begin{figure}
  \centering     
  \includegraphics[angle=0,width=\columnwidth]{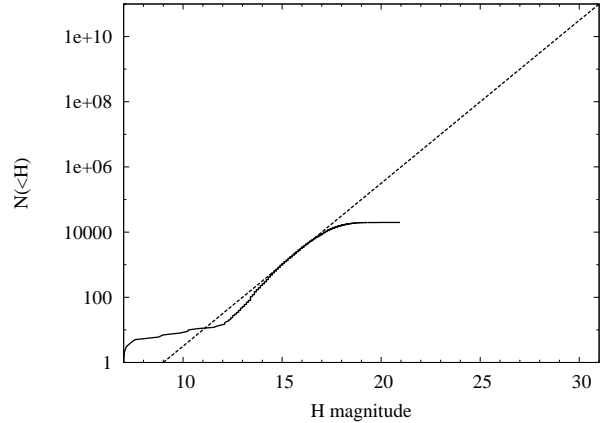}
  \caption{Size frequency distribution (SFD) of the asteroid
    population in the impact region. The SFD is plotted in terms of
    the cumulative number of asteroid as function of their absolute
    magnitude (H).}
   \label{Fig:SFD} 
\end{figure}

Once identified the region of potential impactors, we proceded to
estimate the likelihood and the frequency of impacts, which depend on
the sizes, number density of asteroids, and average impact probability
in the region. The average impact probability in the region was
derived by averaging the intrinsic impact probabilities (one of the
output of our algorithm) of the asteroids in the region; we found
$P_i=3.\, 10^{-19}$ impacts/km$^2$/year.

First of all, we estimated the number of impacts and the size of the
impactors hitting \tc\ during its collisional lifetime assuming a
diameter of 4 m (given by its H value of 30.9 \citep{jenniskens09} and
assuming a geometric visible albedo of 0.05).  The impact probability
per projectile for \tc\ is given by $P_p=R^2 \tau_{coll} P_i$ with $R$
the radius of \tc\ and $\tau_{coll}$ its collisional lifetime
estimated from \cite{marchi06} ($\tau_{coll}=16.2$ Myr). As a
consequence, we find that \tc\ has an impact probability per
projectile of the order of $2.\, 10^{-17}$ during its collisional
lifetime.

To obtain the number of impacts on \tc\ during its collisional
lifetime, we estimated the number of projectiles by calculating the
size frequency distribution (SFD) of the population of the asteroids
in the impact region and extrapolating the corresponding function to
asteroid sizes below the detection limit of current discovery
surveys. Figure~\ref{Fig:SFD} shows the SFD in terms of the cumulative
number of asteroid as function of their absolute magnitude ($H$). We
used $H$ values from the ASTORB file.  In order to extrapolate the SFD
asteroid sizes below the current detection limits, we calculated the
equation of the straight line that best fits the base-10 logarithm of
the known SFD and that is compatible with the size frequency
distribution of the whole Main Belt \citep[see][]{bottke02} (see
Figure~\ref{Fig:SFD}). The line has equation $y=\alpha H + \beta$
where $y$ is the $\log_{10}(N(>H))$ 10-base logarithm of the
cumulative number of asteroids with $H$ smaller or equal than a
threshold value. The best-fit value for $\alpha$ and $\beta$ are $0.5$
and -4.5 respectively, implying about $10^4$ asteroids with $H<17$,
$10^6$ objects with $H<21$, $10^9$ with $H<27$ and $10^{11}$ with
$H<31$, if the linear extrapolation is assumed to be valid.  We can
deduce that for $H<31$, we obtain only $\sim 10^{-6}$ impacts with
\tc.  In the end, 1 impact at v $\leq$ 0.5 km/s during \tc\
collisional lifetime would require $1/P_p$ objects (i.e. $5.\,
10^{16}$ objects).  Using the SFD equation, this means we need a
collision with an object of magnitude $H\sim42$. This correspond to
the size of a pebble of 2.4 cm of diameter. Clearly the contamination
brought by low speed impacts is very small (the volume ratio between
\tc\ and the pebble is 1.1\,$10^{-7}$) and certainly not at the level
of 20-30\% as implied by the number of chondrites found amongst the
ureilites of the \as\ meteorites.

Much more likely, \tc\ was a fragment of a larger asteroid that was
liberated by an impact near the $\nu_6$ resonance few millions years
ago. In fact, we note that (1) the dynamical time to deliver asteroids
from the $\nu_6$ resonance to orbits similar to that of \tc\ is of the
order of 19 Myr \citep[see][for 1999 $RQ_{36}$ coming from the Polana
region]{campins10}; (2) the cosmic ray exposure age of \as\ meteorites
is of 19.5$\pm$2.5 Myr \citep{welten10}; as a consequence, \tc\ spent
most of its life as a single asteroid traveling from the $\nu_6$
resonance to the orbit that brought it to impact the Earth.

It makes sense to investigate what was the contamination brought by low
velocity impacts on the parent body of \tc. The average impact
probability implies that we have $\sim$ $1.4$\,$10^{-9}$
impacts/km$^2$ over the age of the Solar System, which correspond to 1
impact of an asteroid of $H\sim27$ taking into account the SFD of the
objects in the region. The diameter of an asteroid with H=27 is
between 10 and 25 m depending on the albedo. Again the level of
contamination is very little, of the order of 1.\,$10^{-5}$.

When we take into account the necessity of mixing meteorites of at
least three different mineralogies, we conclude that the formation of
\tc\ by means of low-velocity collisions is not a realistic scenario.

\section{Conclusion} 
\label{section_conclusion}

From our study, it seems that the Nysa-Polana family and the
Background at low inclination are good candidates for the origin of
\tcc\ and \as.  First of all, as mentioned in Section
\ref{section_dynamics}, the Nysa-Polana family is located close to the
$\nu_6$ secular resonance which is the favorite route leading to
primitive NEOs and more particularly to asteroid \tcc. Moreover, the
proper inclination of the Nysa-Polana family is very similar to that
of \tc, which should have been maintained during the transfer of \tc\
to the NEOs region through the $\nu_6$ secular resonance.  We also
know, from our algorithm of spectral classification (Section
\ref{section_method}), that the Nysa-Polana family gathers the 3
spectral classes (S, B, and X), which are proposed analogs to \as\
fragments (Section \ref{section_polana}) under the hypothesis that
ureilites are linked to B-class asteroids. More specifically, (1) the
mean spectrum of B-class asteroids of the Nysa-Polana family is
spectrally matched -- at least in the visible -- with available
spectra of ureilitic fragments of Almahata Sitta, (2) considering
space-weathering effects, the mean spectrum of S-class asteroids of
the Nysa-Polana family is compatible with the spectrum of the
H-chondrite fragment $\#25$, (3) a good agreement is found between
X-class asteroids and enstatite chondrites from other meteorite falls
(we remind that enstatite chondrites are part of \as). Of course, a
comparison with enstatite chondrites from \as\ fragments would be very
useful to get a definitive match between the Nysa-Polana family and
\as.  In Section \ref{background}, the same kind of work was performed
for objects of the inner Main Belt coming from the Background at low
inclination ($i<8^\circ$). We concluded that these dispersed asteroids
could also be at the origin of \tc.

In Section \ref{section_formation}, we tried to explain the formation
of \tc\ by low velocity impacts (below $0.5$ km/s) between different
mineralogies in the neighborhood of and within the Nysa-Polana
family. Selecting \tc\ as the target asteroid (d=4 m), we find a
probability per projectile about $10^{-17}$ impacts during its
collisional lifetime (i.e. in 16.2 Myr).  As a consequence, impacts at
low velocity are extremely rare and there is little chance that TC3
was formed by low-velocity impacts in the current asteroid belt. This
implies that the heterogeneous composition of the parent body of \tc\
has to be inherited from a time when the asteroid belt was in a
different dynamical state, most likely in the very early Solar System.
One could think that an asteroid of ureilite composition was
contaminated by impactors of different nature when the asteroid belt
was still massive and dynamically cold, so that mutual collisions were
frequent and at low velocity. However, this view is probably
simplistic. In fact, a body of ureilite composition needs to be formed
in the interior of a large carbonaceous asteroid which underwent
significant thermal alteration \citep{singletarygrove03}. This
asteroid needs to have undergone a collisional disruption to expose
the ureilite material in space. The same is true for the bodies of Hn
composition, with n larger than 3 \citep{gopel93}. But collisional
disruptions require large relative velocities, in contrast with the
view of a dynamically cold belt. Thus, the asteroid belt could not be
overall dynamically cold when the parent body of \tc\ formed.

These considerations suggest that, conversely to what is usually
thought, accretion and collisional erosion had to co-exist for some
time in the asteroid belt. For this to be possible, presumably there
had to be still a significant amount of gas in the system so that,
although large asteroids could be on dynamically excited orbits, the
orbit of their small fragments were rapidly re-circularized by
gas-drag.  Consequently, the mutual relative velocities of these
fragments were small and a new phase of accretion was possible for
them.

We remark that the heterogeneity of \tc\ is not at the microscopic
level; each of the meteorites delivered to the ground are of a
distinct class.  Thus, \tc\ seems to be an agglomeration of
meteorite-sized (i.e. few dm) pebbles of different nature. Pebbles of
this size are strongly coupled with the gas and are extremely
sensitive to pressure gradients. They play the key role in the new
models of planetesimal formations, based on the concentration of
dm-size pebbles in the eddies of a turbulent disk and on the process
of streaming instability \citep{johansen07,johansen09}. These models
of planetesimal formation in a turbulent disk seem a priori to be
particularly favorable to explain the coexistence of collisional
erosion and accretion. Large planetesimals are formed by the
concentration of a large number of pebbles, forming self-gravitating
clumps. Once formed, the orbits of these large planetesimals are
rapidly excited by the stochastic gravitational perturbations exerted
by the turbulent disk \citep{ida08,morby09}. If the threshold for
collisional break-up is achieved, the pebble-size fragments of these
large bodies are re-injected into the game: by being concentrated into
new eddies they can give origin to new large planetesimals and so
forth. Admittedly, quantitative work is needed to support this
scenario; also, other more classical planetesimal formation mechanism
in presence of gas drag \citep{wetherillstewart93, kenyonbromley04,
  weidenschilling11} might explain the coexistence of erosion and
accretion as well.

In this respect, it will be important to understand from the
observational point of view if macroscopic heterogeneity as that of
\tc\ is the exception or the rule among asteroids. \tc\ is the first
object of this kind that has ever been observed, but it is also the
first fall of an asteroid on Earth documented live and for which an
extensive and exhaustive search for meteorites has been done. So, it
might not be as rare as one could be tempted to believe. Indeed, a
second similar case has just been reported \citep{spurny12}. Now that
the possibility for macroscopic heterogeneity is recognized, careful
investigations (also conducted by remote sensing techniques) may
reveal additional interesting cases. Understanding which fraction of
the asteroids are of primary or secondary accretion will be a
fundamental step to constrain the asteroid formation and evolution
models.

\section{Acknowledgements} 
We thank O. Michel and P. Bendjoya for providing us their method of
classification as well as A. Cellino, P. Tanga, M. M\"{u}ller,
H. Campins, B. Carry, and P. Vernazza for helpful discussions.
Programming tools made available to us by the Gaia Data Processing
Analysis Consortium (DPAC) have been used within this work.
J. Gayon-Markt is also grateful to the Centre National d'Etudes
Spatiales (CNES) for financial support.

\end{document}